# DistHash: A robust P2P DHT-based system for replicated objects

**Ciprian Dobre\*. Florin Pop\*, Valentin Cristea\***

\*Computer Science Department, Faculty of Automatic Control and Computes Science,
University POLITEHNICA of Bucharest
(e-mails: {ciprian.dobre, florin.pop, valentin.cristea}@cs.pub.ro)

**Abstract:** Over the Internet today, computing and communications environments are significantly more complex and chaotic than classical distributed systems, lacking any centralized organization or hierarchical control. There has been much interest in emerging Peer-to-Peer (P2P) network overlays because they provide a good substrate for creating large-scale data sharing, content distribution and application-level multicast applications. In this paper we present DistHash, a P2P overlay network designed to share large sets of replicated distributed objects in the context of large-scale highly dynamic infrastructures. We present original solutions to achieve optimal message routing in hop-count and throughput, provide an adequate consistency approach among replicas, as well as provide a fault-tolerant substrate.

## 1. INTRODUCTION

Peer-to-peer (P2P) systems have been widely adopted for communication technologies, distributed system models, applications, platforms, etc. Such systems describe not a particular model, architecture or technology, but rather group a set of concepts and mechanisms for decentralized distributed computing and direct P2P information and data communication.

Currently there has been much interest in P2P network overlays (distributed systems that are not under any hierarchical organization or centralized control) because they provide (to some extent) a long list of features: selection of nearby peers, redundant storage, efficient search/location of data items, data permanence or guarantees, hierarchical naming, security. P2P networks can potentially offer an efficient self-organized, massively scalable and robust routing architecture in the context of wide-area networks, combining fault tolerance, load balancing and explicit notion of locality.

DHT-based systems (Lua, et al, 2004) are an important class of P2P routing infrastructures. They support the rapid development of a wide variety of Internet-scale applications ranging from distributed file and naming systems to application-layer multicast. They also enable scalable, wide-area retrieval of shared information. However, current DHT implementations face several problems. They assume that all peers equally participate in hosting published data objects or their location information, an assumption that can lead to a bottleneck at low-capacity peers. Also, in real-world P2P systems peers join the network according to loose rules, without any prior knowledge of the underlying topology.

In this we present an original DST-based P2P system, DistHash, that optimally shares sets of distributed objects in highly dynamic large scale infrastructures. The system is based on the implementation of a DHT (distributed hash table) in which a) peers do not equally participate in hosting published data objects; b) peers can join or leave the network at any time, without prior knowledge; c) the underlying network infrastructure can be different than the adopted communication scheme. The rest of this paper is organized as follows. In Section 2 we present related work. In Section 3 and 4 we provide an extensive presentation of the concepts and algorithms used by the proposed system. In Section 5 we present several case scenarios and in Section 6 we conclude and present future work.

## 2. RELATED WORK

DHTs (Distributed Hash Tables) are decentralized distributed systems that partition the keys inserted into the hash table among the participating nodes. They usually form a structured overlay network in which each communicating node is connected to a small number of other nodes. Any DHT could be easily turned into a system offering communication services.

The Content Addressable Network (CAN) (Ratnasamy, et al, 2001) is a distributed decentralized P2P infrastructure that provides hash-table functionality on Internet-like scale. CAN is designed to be scalable, fault-tolerant, and self-organizing. The architectural design is a virtual multi-dimensional Cartesian coordinate space on a multi-torus. It routes messages in a d-dimensional space, where each node maintains a routing table with $O(d)$ entries and any node can be reached in $O(dN^{1/d})$ routing hops. Unlike other solutions, the routing table does not grow with the network size, but the number of routing hops increases faster than $\log N$. Still, CAN requires an additional maintenance protocol to periodically remap the identifier space onto nodes.

Pastry (Rowstron& Druschel, 2001) is a scalable, distributed object location and routing scheme based on a self-organizing overlay network of nodes connected to the Internet. Pastry performs application-level routing and object location in a potentially very large overlay network of nodes connected via the Internet. It can support a variety of peer-to-peer applications, including global data storage, data sharing, and group communication and naming. Pastry is completely decentralized, scalable, and self-organizing; it automatically adapts to the arrival, departure and failure of nodes.

Sharing similar properties as Pastry, Tapestry (Zhao, *et al*, 2004) employs decentralized randomness to achieve both load distribution and routing locality. The difference between Pastry and Tapestry is the handling of network locality and data object replication. Tapestry's architecture uses variant of the distributed search technique, with additional mechanisms to provide availability, scalability, and adaptation in the presence of failures and attacks.

The Chord protocol (Stoica, *et al*, 2003) uses consistent hashing to assign keys to its peers. Consistent hashing is designed to let peers enter and leave the network with minimal interruption. It is closely related to both Pastry and Tapestry, but instead of routing towards nodes that share successively longer address prefixes with the destination, Chord forwards messages based on numerical difference with the destination address. Although Chord adapts efficiently as nodes join or leave the system, unlike Pastry and Tapestry, it makes no explicit effort to achieve good network locality.

The Kademlia (Maymounkov& Mazieres, 2002) P2P decentralized overlay network assigns each peer a NodeID in the 160-bit key space, and (key,value) pairs are stored on peers with IDs close to the key. A NodeID-based routing algorithm is used to locate peers near a destination key. One of the key architecture of Kademlia is the use of a novel XOR metric for distance between points in the key space. XOR is symmetric and it allows peers to receive lookup queries from precisely the same distribution of peers contained in their routing tables. Kademlia can send a query to any peer within an interval, allowing it to select routes based on latency or send parallel asynchronous queries. It uses a single routing algorithm throughout the process to locate peers near a particular ID.

The Viceroy (Malkhi, *et al*, 2002) P2P decentralized overlay network is designed to handle the discovery and location of data and resources in a dynamic butterfly fashion. Viceroy employs consistent hashing to distribute data so that it is balanced across the set of servers and resilient to servers joining and leaving the network. It utilizes the DHT to manage the distribution of data among a changing set of servers and allowing peers to contact any server in the network to locate any stored resource by name. Its diameter of the overlay is better than CAN and its degree is better than Chord, Tapestry and Pastry.

## 3. DESIGN OF DISTHASH

The architecture is described in Figure 1. The system is composed of peers (called Agents) and super-peers (called RAgents). All Agents store local replicas used to store distributed objects. RAgents are special peers that, in addition to local replicas, have also meta-data catalogues. A set of Agents is under the control of a central RAgent, those forming a cluster of peers. The RAgents form a complete graph of connections, and control messages are routed among RAgents. Thus, the RAgents connect together several connected clusters of Agents. Such a hierarchical interconnection topology ensures an optimum control scheme and best reflects the real-world phenomenon of large Internet-scale systems. As demonstrated in (Barabasi, *et al*, 2000), networks as diverse as the Internet tend to organize themselves so that most peers have few links while a small number of peers have a large number of links.

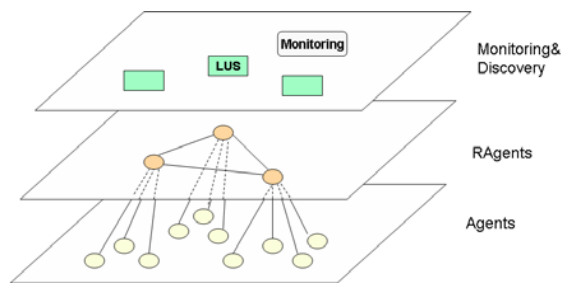

Figure 1. DistHash's architecture.

In addition we assume the existence of several lookup&discovery services located at different locations. These services are used to store information regarding current existing RAgents.

The modality to connect the peers into clusters is based on the geographic position of each peer, as well as a set of condition metrics. The idea is based on our previous work (Dobre, *et al*, 2007), where we presented an algorithm designed to assist the EVO (a P2P videoconference system) peers to dynamically detect the best reflectors to which to connect to. The algorithm considers that the best reflectors are chosen based on their network and geographic location (Network domain, AS domain, Country, Continent), and also based on their current load values, number of currently connected clients, current network traffic.

In our architecture we too consider the existence of several LUSes from where a newly entered Agent reads the current list of RAgents (network address and port on which to connect). It then tries to estimate the best RAgent to which to connect. In this case, the optimum RAgent is chosen based on the network and geographic position. But the newly entered Agent will also request the number of other Agents connected to several of the closest RAgents and will use this to compute a metric such that, in the end, the Agent will connect to one of the closest RAgent having the smaller number of Agents connected. This scheme was previously introduced in case of the EVO P2P videoconferencing system (Legrand, *et al*, 2005).

The way clusters are formed follows an idea borrowed from the way PES (pending event structures) are built. As introduced in (Brown, 1998), Calendar Queue is one of the most advanced PES structure, built upon the idea of dividing the queue into buckets. Similarly, we consider that we have a number of RAgents (equivalent to a bucket) connecting several Agents (length of the bucket or bucketwidth). The determination of the number of Agents connected to a RAgent and of the number of RAgents is crucial to maintaining efficiency. If the number of Agents is too great as compared to the number of RAgents, the meta-data catalogue will grow too big and the operations on it will require a longer time to complete. On the other hand, if the number of Agents is too low as compared to the number of RAgents, the number of control messages required to access information from the system will be higher. This is shown in Figure 2.

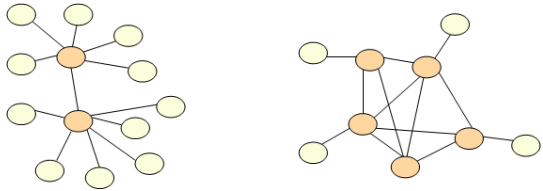

Figure 2. The clustering when there are few RAgents (left) or too many RAgents (right).

The solution given is to allow the number of RAgents increase and decrease correspondingly as the number of Agents grows (new Agents join the system) and shrink (Agents leave the system). This is done on a cluster level. Whenever the number of Agents connected to a RAgent is too high, a newly RAgent is chosen from the Agents (based on a voting algorithm) and the cluster is divided in two, by splitting the remaining Agents between the two RAgents. The meta-data catalogue is also divided between the two RAgents, and the LUS information is updated accordingly. This is best illustrated in Figure 3.

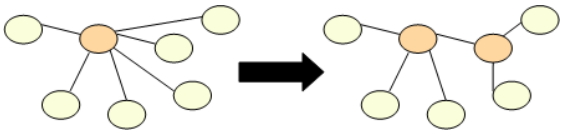

Figure 3. Forming of a new cluster.

On the other hand, whenever the number of Agents connected to a RAgent is too low, the cluster is destroyed by combining the Agents with the Agents of another cluster. The meta-data catalogue is also merged with the catalogue of the RAgent from the second cluster, and the LUS information is updated accordingly. This is best illustrated in Figure 4.

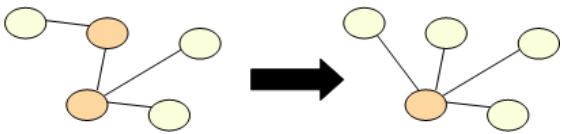

Figure 4. Merging of clusters.

The two operations involve exchange of several control messages to update the current status of clusters, as well as the possible reconnection of several Agents to new RAgents and RAgents with other RAgents. In order to minimize the number of costly operations we consider thresholds for what too few or too many agents mean to be the values 5 and 10.000. These values are based on the observations from (Brown, 1988).

Inside a cluster peers connect as represented in Figure 5. Because RAgent stores the meta-data catalogue, it must not represent a central-point-of-failure for the entire cluster. For this reason a specially designated Agent will also act as a secondary backup replica for the RAgent. In case of failure the secondary Agent will take over the original RAgent. All updates of the meta-data catalogue will immediately also propagate to the secondary RAgent replica.

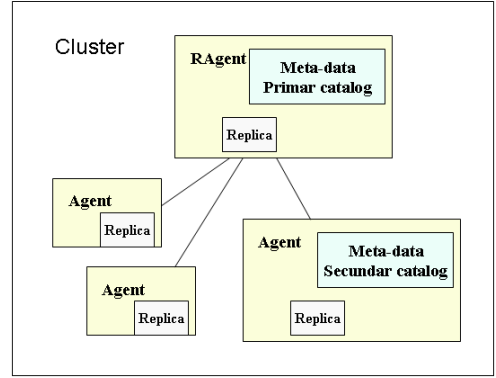

Figure 5. Roles of peers inside a cluster.

## 4. DATA MANAGEMENT

As presented, each Agent maintains a copy of several replicated objects. These are Java serializable objects extending DistData (see Figure 6). The objects can represent anything, from basic data types to blobs or even data files.

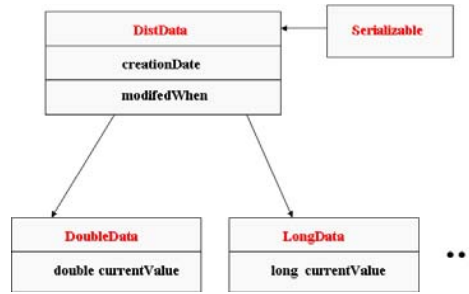

Figure 6. Distributed objects.

In order to ensure data consistency among replicas we adopt a simple strategy and require that each object has an owner (an Agent that has the right to modify the current state of a particular object). The ownership, as well as information regarding current objects existing inside a cluster, are maintained by the meta-data catalogue (Figure 7). \

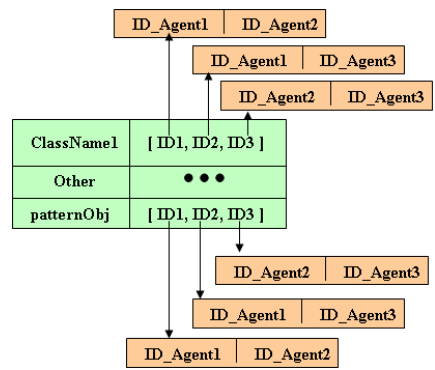

Figure 7. Information stored in the meta-data catalogue.

The meta-data catalogue is organized in the form of a hash table, where the keys represent specific information regarding objects (such as name of the Java class, specific access patterns or other information the user creating a new object declares as being of interest). The values represent linked lists of (key, value) mappings. The key in the linked list represents an object identifier matching a particular pattern key (ID1, ID2, etc.). An object identifier represents a hash uniquely identifying a particular object. Each value also points to a linked list of Agent identifiers (ID_Agent1, ID_Agent2, etc). The first identifier always points to the Agent currently having the ownership over a specific object.

Inside a replica, the objects are stored in a hashtable, where the key represents the object identifier and the value represents the object itself. In order to access a particular object a request would then travel to a RAgent, and from there to a particular Agent. The search operation on such a structure involved several stages, as presented in Figure 8.

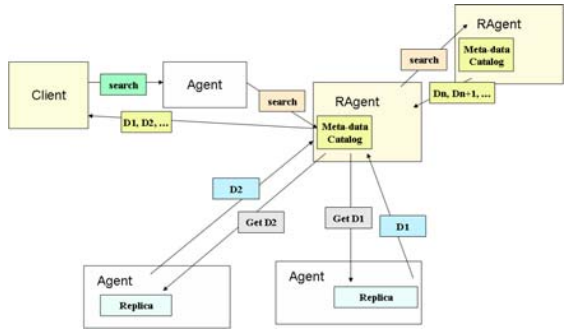

Figure 8. The search operation.

First a client initiating a search would contact an Agent (the LUSes are secured and no client has direct access to internal information regarding the list of RAgents). The Agent would then initiate the search process in the name of the Client. It will contact one RAgent and send the current search criterion. Inside a RAgent the search translated into finding the key in the meta-data catalogue corresponding to the search criterion. This operation requires $M$ steps, where $M$ represents the number of keys in the meta-data catalogue. This could be improved, for example, by considering separate catalogue for exact information (class names) and for search patterns. Searching for the objects corresponding to a particular class name could then require only $log\ M$ operations (using the Java mapping of hash keys for strings corresponding to the string values).

After finding particular entries corresponding to the search criteria, the search involves obtaining the unique IDs of corresponding objects, as well as an Agent ID owning a particular object. This requires $2P$ steps, where $P$ is the number of objects matching the search criteria. Then the RAgent will contact each Agent and request a copy of the current object having a particular id. This involves again $2P$ messages being sent and received (in fact, the RAgent will not send one request for each object detain by a particular Agent, but will group the requests into one single one and send it once – the actual number of exchanged messages is smaller then $2P$). An agent will get a particular object corresponding to a particular id in $log\ L$ steps (a hash search), where $L$ represents the number of objects currently possessed by an Agent. In the end, the data is collected and sent back to the Agent that initiated the search operation, and back to the Client.

Among RAgent the search operation involves the following operations. A RAgent will send a request with a particular search criterion. Each RAgent will respond back with the list of objects matching the criteria, together with their corresponding object identifiers. The first RAgent will then merge the received lists of objects, together with its own list, such that, in the end, no two objects having the same identifiers will be returned in the search. In the end, considering $R$ RAgents, the total number of steps involved by the search is given by $S = R * M * (4P+log\ L)$.

The insert operation involves several steps. First, a client willing to add a new object into the system would contact an Agent. The Agent will then send the insert request to the RAgent to which is connected. The RAgent then selects two Agents that will further keep two replicas of the object stored locally. The RAgent updates the meta-data catalogue with the information regarding the new object, selects one Agent as being the owner of the object and then sends the replicated objects to the two newly selected Agents. The selection is done such that to obtain a balancing of the number of objects detain by each Agent. This is accomplished by carefully maintaining in the memory of the RAgent a list of the number of objects detained by each of the connected Agents and then selecting two Agents that previously kept too few objects. This approach guarantees a balance the load of the Agents and also, as detailed later, ensures a highly fault tolerant environment.

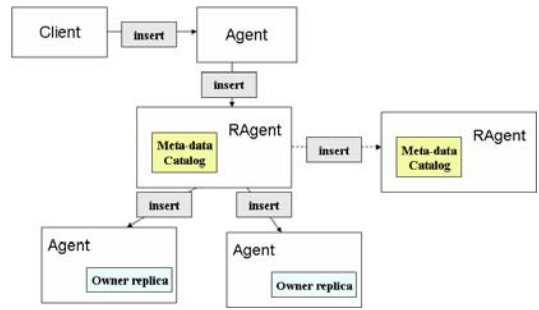

Figure 9. The insert operation.

Assuming an ideal equal split of the number of objects between all Agents in the system and considering the only search criteria is the class names of the objects, $M = B / R$, where $B$ represents the number of objects at a particular moment kept in the entire system (the objects would be equally divided among clusters). The number of objects matching a particular class name would then be $P = 1$. Also, the number of objects kept by an Agent is given by $L = (2B / R) / (N / R)$, meaning the total number of objects kept inside a cluster $(2B / R)$ divided by the number of Agents in a cluster $(N / R)$, where N represents the total number of Agents. In the end, $S=B*(4+log2B/N)$. This means that, as the number of objects increases, the complexity of the search operation increases, but in the same time when the number of Agents increase the complexity of the search operation decreases.

In reality, using this approach the objects could in fact accumulate more around a single cluster. To cope with situation the RAgent could also delegate the insert operation, if the size of the local catalogue grows too big, to other RAgents. This is best illustrated in Figure 9.

Also, the search operation can be optimized considering the following approach. A special query search message is implemented with the syntax of return the first object matching a particular criterion. Then, when a RAgent receives two many searches for a particular object, located in another cluster, it could decide to move the object inside the local cluster. In this way, the search does not travel anymore all RAgents and is faster served back to the client. Also, the read operation can be performed on any Agent having a replica of a particular object. A special request search syntax can return back the address of an Agent having a particular replica object. The client can then initiate successive read operations on that particular replica, directly accessing one particular Agent. This greatly reduces the number of steps involved and offers greater flexibility to the client (application developers using the system).

The update operation is an asynchronous operation, where an Agent initiates an update operation on a particular object. The RAgent keeps a lock on the owner of the object and then, asynchronously, updates all active replicas of that objects. As soon as the RAgent receives the request and performs the lock it sends back a notification message that the update operation is in progress. This approach ensures consistency among the replicas (no two updates would happen simultaneous on different replicas because only one owner of a particular object exists in the system).

## 5. FAULT TOLERANCE MANAGEMENT

According to the previously described scheme, faults are eventually detected in the system. For example, when a client initiates an update operation the RAgent could discover that a particular Agent is no longer active in the system. This is the reactive approach. Also, we consider a proactive fault detection approach, Agents and RAgents periodically sending (where period is high enough not to introduce too much communication overhead in the system) heartbeat messages among them.

Whenever an Agent fails a process of reallocation objects is initiated by the RAgent (Figure 10). First, all objects previously owned by the crashed Agent will now be owned by the surviving Agents still having replicas of that objects. In this manner we ensure that Agents running for a long time will eventually become owners of more and more objects and the number of messages needed to change ownership would, therefore, in long term decrease. Then, the RAgent selects one surviving Agent to be the new keeper of the secondary replica of an object previously detain by the crashed Agent. The algorithm used to select the Agent considers a balancing of the number of objects among surviving Agents at all times.

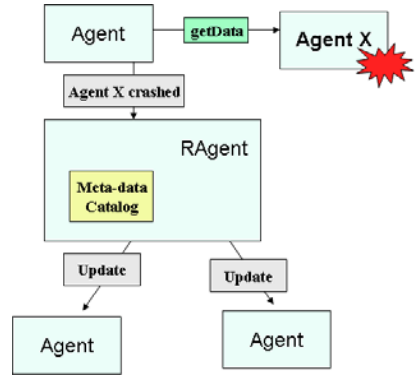

Figure 10. A crash scenario of an Agent.

The second crash scenario involves the disappearance of the RAgent itself. In this case the secondary RAgent replica takes its place (updates the entry in LUSes, and connects to all other RAgents) and, among remaining Agents, one is selected (based on a voting procedure) to become the new RAgent replica (see Figure 11)..

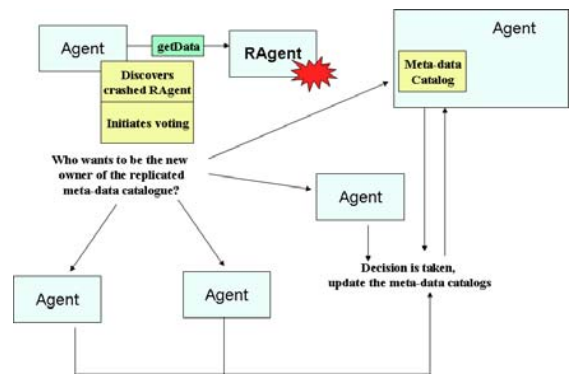

Figure 11. A crash scenario of an RAgent.

Of course, the voting procedure should be consistent, such that if two agents observe at closely the same time the disappearing of the RAgent they both initiate the voting procedure, but in the Agent selected to be the new RAgent replica is the same.

In the event of a transient failure (an Agent or RAgent temporarily leaving the system and then entering back at a later time) the RAgent (newly or one from another cluster) detects the situation and informs the newly entered peer of the changed. In this way, we can cope with both transient and permanent failures of nodes in the system. Also, the system is highly resilient to faults as allows almost a very large number of nodes to fail without loosing any information (of course, if both Agents having the replicas of the same object fail in the same time the object is last; however, if the two Agents fails at different moments of time – even if they are close – there is a high probability that a new Agent will have already take the role of the first crashed Agent). The information is always rapidly replicated so that to offer a maximum degree of fault tolerance level to clients.

## 6. CONCLUSIONS

In this paper we presented DistHash, a P2P system that emulate the functionality of a DHT. DistHash is a P2P overlay network designed to share large sets of replicated distributed objects in the context of large-scale highly dynamic infrastructures. We presented the original adopted solutions to achieve optimal message routing in hop-count and throughput, provide an adequate consistency approach among replicas, as well as provide a fault-tolerant substrate. All this are

The system provides a middleware level functionality to users needing to use a shared memory concept system to be accessed by distributed application running on an Internet-like large-scale infrastructure. The problems with previous approaches were their lack of scalability, adaptation to the dynamic nature of peers (joining or leaving the system at any time), and they did not consider the influence of the underlying existing network infrastructure.

DishHash is highly robust in the face of failures happening in various points of the system, as well as peers highly dynamically entering or leaving the system. It also offers a balanced approach to the way replicated objects are kept inside the system. The operations are optimized in terms of steps required, as well as number of messages exchanged between peers. For that we consider a hierarchical approach in which we organize peers in clusters. Inside each cluster the functionality is almost autonomous so that the total number of peers would not influence the performance of the system. This means that DistHash also offers scalability to end-users.

The system is currently well under implementation and we plan to test its performance compared to other existing solutions. We also plan to adopt several improvements in the routing, consistency, searching and fault-detection algorithms so that to improve even more its performance. For that, we plan to integrate DistHash also with a monitoring instruments and use the monitoring information to take decision regarding the replicated objects in real-time (Dobre, *et al*, 2007).